\documentclass[a4paper,reqno,12pt,draft]{article}
\usepackage{amssymb,euscript,bbold}
\newcommand{\be}{\begin{equation}}
\newcommand{\ee}{\end{equation}}
\newcommand{\ba}{\begin{eqnarray}}
\newcommand{\ea}{\end{eqnarray}}

\newcommand{\ft}{\footnote}

\begin{document}
\input{epsf}

\begin{flushright}
\end{flushright}
\begin{flushright}
\end{flushright}
\begin{center}
\Large{\sc Observations on the Space of Four Dimensional
String and $M$ theory Vacua\ft{Summary of a talk given
at the occasion of the retirement of David Olive, University of
Wales, Swansea, March 2004}}\\
\bigskip
\bigskip
\large{\sc B.S. Acharya}
{\renewcommand{\thefootnote}{}
\footnotetext{bacharya@ictp.trieste.it}}\\
\bigskip\large
{\sf Abdus Salam\\ International Centre for Theoretical 
Physics,\\
Strada Costiera 11\\ 34100 Trieste. Italy.}
\end{center}
\bigskip
\begin{center}
{\bf {\sc Abstract}}
The space of four dimensional string and $M$ theory vacua
with non-Abelian
gauge symmetry, chiral fermions and unbroken supersymmetry beyond
the electroweak scale appears to be a disconnected space whose different
components represent distinct universality classes of vacua.
Calculating statistical distributions of physical observables a la
Douglas therefore requires that the distinct components are carefully
accounted for. We highlight some classes of vacua which deserve
further study and suggest an argument which may serve to rule
out vacua which are small perturbations of supersymmetric $AdS_4$.

\end{center}


\newpage


String/$M$ theory appears to have {\it many} vacua.
Many known examples of these vacua do not correspond 
to universes with
just four macroscopic spacetime dimensions, an $SU(3)\times SU(2)
\times U(1)$ gauge symmetry at the electroweak scale, 
chiral fermions  or many other 
observed properties of our universe.
For example, eleven dimensional Minkowski spacetime,
the $E_8 \times E_8$ 
heterotic string compactified to six dimensions on a
$K3$ surface, or Type IIB string theory on $AdS_5 \times S^5$
with $N$ units of flux are strongly believed - because of
various strong-weak coupling dualities - to be {\it exact} vacua.

This is {\it not} a problem for the theory if there {\it does} 
exist
a suitably stable ``correct vacuum''. The main problem for theory
is to demonstrate the existence of such a vacuum.
Indeed, there is no
good reason that we know of that implies 
one vacuum in a quantum theory
of gravity should be preferred over another and the proliferation
of incorrect string vacua suggests that we may never have such a
reason. However, a proper understanding of cosmology in string/$M$
theory may, ultimately, lead to selection rules \cite{douglas1}.  

Given the lack of any known selection rules,
we therefore need to understand the space of realistic string/$M$
theory vacua. To make progress with such an understanding clearly
requires  choices. 
For example, in field theory one can conceive
of many extensions of the standard model beyond 
the electroweak
scale. This is reflected in string theory also:
there may be
vacua with classically broken supersymmetry at the 
compactification
scale or vacua with supersymmetry unbroken at high energies. 
Similarly,
some stable vacua may have large extra dimensions others small
extra dimensions. 

Furthermore, we do not have a complete list of all vacua
and even if we did
would presumably not have time to wade through the list and
single out all of those that agree
with the standard model at the electroweak scale and below.
In view of these issues,
we will consider
vacua which have some basic properties that we will deem
essential to all realistic vacua. This is the
practical viewpoint proposed in \cite{douglas1}.

We will make the following choices: four macroscopic spacetime
dimensions, non-Abelian gauge symmetry (with rank $\geq$ four),
chiral fermions and unbroken supersymmetry at some scale
above $M_{ew}$. 

{\it Remarks.} 1.The first three choices are 
justified by experiment.
2. 
Whilst the existence of supersymmetry is purely conjectural, in
addition to the standard reasons for including it, there is a
more practical reason: string vacua with classically broken
supersymmetry are poorly understood and may even be unstable.

We will denote the space of string/$M$ theory vacua with
these properties as ${\cal M}$. At the present time we have no
idea how big ${\cal M}$ is. Of course, the set of known
examples of 
string/$M$ theory vacua in 
${\cal M}$ provide us with a {\it lower bound} 
on the number of points it contains. One aim of the
present paper is to present evidence that ${\cal M}$ is
a disconnected space - disconnected in a very physical sense;
namely that there exist points in ${\cal M}$ whose physics is of
a different universality class to other points in ${\cal M}$.
A second aim is to suggest that the space is ``larger'' than
our currently known examples would tend to suggest.

A third point is that the discussion here serves to put the
recent work of Douglas et. al. \cite{douglas1,etal}
into a broader context: one needs to know what
fraction of ${\cal M}$ the known vacua represent before
one can say with confidence that a physical property is
statistically favoured. We will return to this point later.

What do we know about
vacua in ${\cal M}$?
There are many examples of vacua in ${\cal M}$.
A more or less complete list of {\it known} points in ${\cal M}$
can be classified as follows, in no particular order

\bigskip
\bigskip

\noindent
a. $E_8 \times E_8$ or $SO(32)$ heterotic strings on 
Calabi-Yau threefolds, $Z_{CY}$\\
b. $M$ theory on $Z_{CY} \times {S^1 \over {\bf Z_2}}$\\
c. Type IIA orientifolds on $Z_{CY}$ with D6-branes\\
d. Type IIB orientifolds on $Z_{CY}$ with D(odd)-branes\\
e. $F$-theory on Calabi-Yau 4-folds.\\
f. $M$ theory on $G_2$-holonomy manifolds with singularities.\\
g. Non-geometric $CFT$'s where the extra dimensions are 
represented by a $SCFT$ without a geometric limit.\\
h. Freund-Rubin compactifications of $M$ theory on singular
Einstein manifolds with $R > 0$.
\bigskip
\bigskip

{\it Remarks.} 1. Vacua in classes $[a],[c],[d]$ can be defined
perturbatively with a world-sheet super conformal field theory 
($SCFT$) and
as such are expected to share many properties with those in class
$[g]$. 2. Under certain conditions, one can add fluxes to some of the
vacua listed here \cite{flux}. 
By definition, this should be taken implicitly.
For example, vacua in class $[e]$ means all $F$-theory vacua with
supersymmetric fluxes. 
\bigskip

{\it Dualities Between Classes.}

Vacua from a given class {\it may} be dual to vacua in another
class. In fact, vacua from classes $[a]$ through to $[f]$ can be
related by dualities. For example, all of the $[c]$ vacua are
also $[f]$ vacua; many of the $[d]$ vacua are limits of $[e]$
vacua; when the $G_2$-holonomy manifolds in $[f]$ vacua are
$K3$-fibered, they are dual to vacua of class $[a]$. So, for
vacua from the first six classes listed, non-perturbative 
equivalences are known to occur. As such, many of these vacua
share many similar properties. For example, the
classical four dimensional vacuum is flat
spacetime.

More importantly, there are vacua in this list which are
{\it apparently not} dual to other vacua in this list.

For example, if the Calabi-Yau fourfolds in $[e]$ or the
$G_2$-holonomy manifolds in $[f]$ are {\it not} $K3$-fibered
then they are apparently not dual to the heterotic string
on a Calabi-Yau threefold.

In fact if we {\it assume} 
that all dualities between string/$M$ theory
vacua originate from {\it fundamental} known dualities between 
vacua with $16$ or more supercharges, then it is clear
that there exist points in ${\cal M}$ not dual to other points
in ${\cal M}$.

This already indicates the disconnectedness of ${\cal M}$.
An important assumption of \cite{douglas1} is that
vacua of type $[d]$ represent a reasonable enough fraction
of vacua in ${\cal M}$ so that statistical analyses of such
vacua give a reasonable picture of the statistics of ${\cal M}$
itself. However if, as we have suggested, vacua exist
which are {\it not} dual to $[d]$ vacua, we need to have a
good idea about how many such vacua there are. For vacua from
classes $[a]$ through to $[f]$ this amounts to knowing for example,
how many Calabi-Yau fourfolds or $G_2$-holonomy manifolds there
are which are not $K3$-fibered.
In the case
of non-dual Calabi-Yau vacua, this might not represent a problem
for such analyses, since it seems reasonable to suppose that 
statistics based on a single Calabi-Yau will not vary 
considerably from Calabi-Yau to Calabi-Yau. However, non-dual
$G_2$-holonomy vacua may have different statistics to
$[d]$ vacua.

Let us also note that, dualities aside, very distinct universality
classes could also be contained within a single type of vacuum.
Consider for instance the $G_2$-holonomy vacua described in our paper
in \cite{flux}. Properties of such vacua depend upon a 
topological invariant $c_{2}$, the imaginary part of a complex
Chern-Simons invariant. If $c_2$ is large, these vacua have large
extra dimensions, if it is small they have small extra dimensions.
We also remind the reader that constructing $G_2$-holonomy manifolds
is technically difficult, so at present
we have no idea if there are more vacua 
large $c_{2}$ vacua than those with small $c_2$.

We will now go on to discuss vacua of class
$[h]$ and describe several features of these which suggest that
they are in a completely
different universality class from the rest. We also emphasise
that there are large numbers of $[h]$ vacua. If, as we suggest,
these vacua are truly distinct from the other types, then
it is crucial for a statistical analysis of the physics of
${\cal M}$ to compare $[h]$ vacua with, say, $[c]$ vacua.

\newpage

{\it Freund-Rubin Vacua}
\bigskip

Thus far, we have only discussed vacua in classes $[a]-[g]$.
A Freund-Rubin vacuum 
\cite{FR} can be described classically as a solution
to d=11 supergravity in which the extra dimensions form a compact
Einstein manifold $(X, g_7 )$ 
with positive scalar curvature and four dimensional
spacetime is anti de Sitter. The background is supersymmetric
if $(X, g_7)$ admits a Killing spinor. Quantum mechanically
the Freund-Rubin vacuum might be defined as the three-dimensional
conformal field theory residing on the world volume of a $N$
$M$2-branes at the tip of a cone with base $X$ \cite{malda,cone}.

In a Freund-Rubin vacuum, non-Abelian gauge symmetry can emerge
from one of two sources:
a) isometries of $g_7$ or b) from $ADE$-singularities
supported along a 3-manifold $Q$ $\subset X$, analagously
to the case of $G_2$-holonomy manifolds \cite{adeg2}.

In order for a supersymmetric Freund-Rubin vacuum with non-Abelian
gauge symmetry to represent a point in ${\cal M}$ it must also
admit chiral fermions.

In \cite{adhl} we demonstrated that such points exist.
The mechanism for chiral fermions is 
identical to that of the $G_2$-holonomy vacua \cite{chiral}:
conical singularities in $X$ of various types.

The Freund-Rubin vacua in ${\cal M}$ which were studied
in \cite{adhl} have various properties which suggest that
they are generally {\it not} 
dual to any of the vacua in ${\cal M}$
of types $[a]$ to $[g]$. 

The vacua discussed in \cite{adhl} roughly come in a two
parameter family labelled by two integers $(N,k)$.
$N$ is the unit of Freund-Rubin flux, which is dual to the
number of $M$2-branes characterising the holographically
dual conformal field theory. $N$ controls the four-dimensional
cosmological constant, which in Einstein frame is of
order
\be
\lambda \sim {m_p^4 \over N^{3 \over 2}}
\ee

This formula demonstrates that $\lambda$ can be parametrically
small in Planck units. This is also true for the supersymmetric
$AdS$ vacua discussed in \cite{flux}. 

$k$ is proportional to the rank of the gauge group, which in the
examples considered in \cite{adhl} is $SU(k)^3$.

Both $N$ and $k$ are free parameters of the supergravity solution
and can, at least classically, take any positive integral value.

Vacua of types $[a]$ - $[g]$ have a gauge group whose rank
is typically not that large eg less than 100. 
Consider a Freund-Rubin vacuum with,
say, $k = 10^8$. It is 
difficult to imagine that it could be dual
to, say a Calabi-Yau compactification with branes. Of course, if the
gauge group were confined at low energies, then perhaps $k$ is
dual to a flux quantum number in Calabi-Yau compactification.
But the gauge group need not be confined in general.

Another aspect of the vacua discussed in \cite{adhl} is that
the supergravity solution is extremely simple compared to
Calabi-Yau or $G_2$-holonomy vacua. This is because one can
more or less {\it explicitly} write the spacetime metric whereas
this is much more difficult for compact, simply
connected Ricci flat manifolds. This is reflected by the fact
that the 7-manifold $X$ of extra dimensions is much simpler 
topologically than a Calabi-Yau or $G_2$-holonomy space.
For instance, the examples discussed in \cite{adhl} are
manifolds fibered by a circle over $S^3 \times S^3$ and the 
circle degenerates in particular ways along spherical submanifolds
of $S^3 \times S^3$ (these are the places at which $X$ has
singularities supporting gauge fields and chiral fermions).
These examples are almost certainly not $K3$-fibered lending
further evidence, albeit mathematical, to the proposition that
they are not dual to other classes of vacua.

\bigskip

{\it Other Classes of Vacua?}

Thus far we have only considered the known types of vacua
in ${\cal M}$ and have offered some arguments that up to
non-perturbative dualities there are many distinct universality
classes. In our opinion, it is only a matter of time before
new vacua in ${\cal M}$ are found. This is just a technical
problem of generally solving the conditions for unbroken 
supersymmetry.

\bigskip

{\it Cutting} ${\cal M}$ {\it down to Realistic Vacua.}
\bigskip

Ultimately we would like to consider only those vacua
in ${\cal M}$ which agree with the standard model at low
energies. Again we face the difficulty of not knowing all
vacua. So we will have to choose questions which are more
general. We will begin by discussing a simple physical objection
to Freund-Rubin vacua. Then we will point out that this 
objection in fact quite general and applies to many vacua 
in ${\cal M}$ including
some of those discussed in \cite{flux}. 
The argument might be used to rule out certain
vacua which are small perturbations of supersymmetric $AdS_4$.
We will emphasise that until
we understand supersymmetry breaking and other corrections to the
theory that these arguments are not strong enough to rule out
more interesting vacua in ${\cal M}$.

The objection to vacua of type $[h]$ is this. 
The classical solution is such that the cosmological
constant of anti de Sitter space is of the same order of 
magnitude (but with opposite sign) as that of the 7-manifold
of extra dimensions. This implies that the masses of Kaluza-Klein
modes are of the same order as the fluctuations of the graviton
in the $AdS_4$. Then, if the $AdS_4$ were large enough to
contain our universe the Kaluza-Klein modes would be too light
to have escaped detection today. However, this argument has two
flaws.

Firstly, it is possible that the radii of the extra dimensions
be parametrically smaller than the radius of $AdS_4$ 
as was demonstrated
in \cite{adhl}. In this case there is a mass gap between $m_{kk}$
and $m_{AdS}$.
However, in practice it is probably impossible
to produce an Einstein manifold whose scalar curvature is
equal to that of a round 7-sphere but whose volume is 
exponentionally smaller which is what is required for
a Freund-Rubin vacuum with $\lambda \sim 10^{-120}m_p^4$
and large enough Kaluza-Klein masses.  

The second flaw is simply that the argument is based on the
behaviour of the classical solution. Quantum corrections, which
are poorly understood in this context, or even
supersymmetry breaking classical corrections might give a vacuum
which is compatible with a large universe and suitably massive
Kaluza-Klein modes. An example of the latter might be obtained
by additional fluxes through $X$, raising the energy of the 
vacuum.

In any case, these physical arguments against Freund-Rubin 
vacua are actually symptomatic of many vacua in ${\cal M}$.
The reason for this is that in a four dimensional supergravity
theory with superpotential $W$ and Kahler potential $K$,
unbroken supersymmetry requires that $\partial_i W + \partial_i K
W = 0$, where the derivative is with respect to all fields.
If the theory contains $p$ fields this is $p$ equations for
$p$ fields. Typically, with ${\cal N} =1$ supersymmetry the
superpotential will be non-zero and hence we expect many 
supersymmetric vacua to exist. Since $W$ will typically
be non-zero
in the vacuum\ft{setting it to zero is an additional condition,
generically incompatible with the first $p$ constraints.}
we have a supersymmetric $AdS_4$ vacuum in which
$\lambda = -3e^K |W|^2$ in Planck units. In vacua with
fluxes and/or quantum corrections accounted for, this can give
rise to $AdS$ vacua whose cosmological constant is much less than
one in Planck units \cite{flux}. 
However, if one considers
such examples in which $|\lambda_{AdS} | \sim 10^{-120} m_p^4 $
then typically the masses of Kaluza-Klein modes are too small.
This is simply because the potential in Einstein frame decreases
at large volume. This is due to the $e^K$ factor being small
\ft{K is quite generally of the form $-ln{Vol(X)}^a$ for
positive $a$}.

An exception to this argument could be
provided by vacua in which the supersymmetric critical point
exists because of  instanton contributions to the superpotential.
If $W$ is consequently very small in the vacuum - a
condition which does
not require particularly large volume - then one would have
cosmologically large $AdS$ space with massive enough Kaluza-Klein
modes. Supersymmetric KKLT vacua, examples of which have
recently been constructed \cite{race}, or $G_2$-holonomy
vacua with fluxes and instanton contributions included, would
be natural places to look for such examples.

So whilst this argument might serve
to rule out vacua which are small perturbations of 
supersymmetric $AdS$,
small superpotentials and vacua in which high 
degree cancellations
to the supersymmetric vacuum energy occur can not be falsified
by it.

In any case, 
supersymmetry breaking and other corrections to the potential
clearly need to be understood better before we can safely
disregard significant numbers of points in ${\cal M}$.

In summary: there appear to be a large number of four dimensional
string and $M$ theory vacua with non-Abelian gauge groups,
chiral fermions and unbroken supersymmetry above $M_{ew}$.
Among these are distinct universality classes whose
contributions to the distributions of physical observables
must be carefully accounted for in the proposal of 
\cite{douglas1}.
Non-$K3$-fibered $G_2$-manifolds
and Freund-Rubin vacua are two notable classes which deserve 
further study. From general supergravity considerations there
are a significant number of supersymmetric $AdS_4$ vacua.
Vacua which are a small perturbation of these in which
the Kahler potential is large and negative 
and the superpotential is not small
might be ruled out on the grounds that the 
extra dimensions are too large.

\bigskip
\large
{\sf Acknowledgements.}
\normalsize
Many of these observations were presented at the University
of Wales, Swansea on the occasion of the retirement of
David Olive in March 2004. We would like to thank
Frederik Denef and Mike Douglas for many discussions.

\end{document}